# Inelastic Light Scattering in the Vicinity of a Single-Atom Quantum Point Contact in a Plasmonic Picocavity

Shuyi Liu[1,†,‡], Franco P. Bonafe[2,†], Heiko Appel[2], Angel Rubio[2,3], Martin Wolf[1], Takashi Kumagai[1,4]*

[1]*Department of Physical Chemistry, Fritz-Haber Institute of the Max-Planck Society, Faradayweg 4-6, 14195 Berlin, Germany.*

[2]*MPI for Structure and Dynamics of Matter, Luruper Chaussee 149, 22761 Hamburg, Germany.*

[3]*Center for Computational Quantum Physics (CCQ), Flatiron Institute, 162 Fifth Avenue, New York NY 10010, USA.*

[4]*Center for Mesoscopic Sciences, Institute for Molecular Science, Okazaki 444-8585, Japan.*

[†]These authors equally contributed to this work.

[‡]Current Address: *Center for Mesoscopic Sciences, Institute for Molecular Science, Okazaki 444-8585, Japan.*

*Corresponding author: kuma@ims.ac.jp

**Abstract**

Electromagnetic fields can be confined in the presence of metal nanostructures. Recently, sub-nanometer scale confinement has been demonstrated to occur at atomic protrusions on plasmonic nanostructures. Such an extreme field may dominate atomic-scale light–matter interactions in "picocavities". However, it remains to be elucidated how atomic-level structures and electron transport affect plasmonic properties of a picocavity. Here, using low-temperature optical scanning tunneling microscopy (STM), we investigate inelastic light scattering (ILS) in the vicinity of a single-atom quantum point contact (QPC). A vibration mode localized at the single Ag adatom on the Ag(111) surface is resolved in the ILS spectrum, resulting from tip-enhanced Raman scattering (TERS) by the atomically-confined plasmonic field in the STM junction. Furthermore, we trace how TERS from the single adatom evolves as a function of the gap distance. The exceptional stability of the low-temperature STM allows to examine distinctly different electron transport regimes of the picocavity, namely in the tunneling and quantum point contact (QPC) regimes. This measurement shows that the vibration mode localized at the adatom and its TERS intensity exhibits a sharp change upon the QPC formation, indicating that the atomic-level structure has a crucial impact on the plasmonic properties. To gain microscopic insights into picocavity optomechanics, we scrutinize the structure and plasmonic field in the STM junction using time-dependent density functional theory. The simulations reveal that atomic-scale structural relaxation at the single-atom QPC results in a discrete change of the plasmonic field strength, volume, and distribution as well as the vibration mode localized at the single atom. These findings give a qualitative explanation for the experimental observations. Furthermore, we demonstrate that strong ILS is a characteristic feature of QPC by continuously forming, breaking, and reforming the atomic contact, and how the plasmonic resonance evolves throughout the non-tunneling, tunneling, and QPC regimes.

Metal nanostructures can confine light to deep sub-wavelength scales and enhance the field strength through excitation of localized surface plasmon resonance (LSPR), whereby the optical properties can be tailored by size, shape and material of nanostructures.[1] LSPR excitation leads to strong light–matter interaction that can be used for highly sensitive optical spectroscopy and efficient photochemical energy/material conversion.[2] Recent studies demonstrated that atomic-scale light confinement can be achieved in the presence of atomic-scale structures (protrusions).[3] Such extreme confinement will dominate the light–matter interactions in a sub-nanometer cavity between metal nanostructures, called picocavity,[4] which enables ultrasensitive optical spectroscopy at atomic scales and extremely strong light–matter coupling.[5] First-principles simulations have predicted a potential impact of atomic-level structures on plasmonic properties of metal nanoparticles,[6,7,8] thus atomic-scale structures may be utilized to engineer optical properties. Indeed, a previous study proposed an optoelectronic device whose property relies on a single atom located in a plasmonic junction.[9] Also, a potential role of atomic structures in light scattering by plasmonic nanostructures was proposed already in the early days of surface-enhanced Raman spectroscopy in 1980s.[10,11,12,13] The atomic-scale optical properties are determined by the complex interplay between LSPR, its coupling to atomic-level structures, motion of atoms, and electron transport properties at atomic contacts.[8,14,15] Therefore, the investigation of atomic-scale optomechanics requires experimental and theoretical methods capable of observing atomic-level structures and their optical properties.

The underlying physical mechanism of extreme light confinement is explained by extra electromagnetic field localization at atomic-scale protrusions, which is superimposed on the plasmonic mode of a metal nanostructure that couples to far-field radiation.[16] Such a non-resonant lightning rod effect is susceptible to the atomic-level structure and its local environment. In addition, plasmonic properties in a picocavity are also affected by quantum mechanical effects,[17] such as electron tunneling and nonlocal screening resulting from spill-out of electrons outside metal surfaces and a finite spatial profile of plasmon-induced charges, respectively.[18,19] These quantum mechanical effects are highly sensitive to the cavity size since electron tunneling depends exponentially on the distance. The conductivity in the plasmonic cavity alters its resonance frequency, local field enhancement, and mode volume.[6,8,16] Although a contribution of atomic-level structures and quantum mechanical effects to plasmonic properties has been extensively investigated by first-principles calculations,[6,8,16] it remains a challenge to experimentally study the microscopic mechanisms.

To validate the theoretical predictions, it is desirable to perform experiments using well-defined model systems that can examine a correlation between plasmonic properties, electron transport, atomic-level geometric and electronic structures. However, accurate experiments in picocavities are typically hampered by instability of atomic-scale structures caused by random movement (diffusion) of metal atoms.[20] Low-temperature experiments can circumvent thermal fluctuation,[4] and the combination with scanning tunneling microscopy (STM) allows to directly characterize the atomic structure and to regulate the cavity size at sub-ångström precision to monitor the plasmonic properties.[15, 21]

The atomically-confined plasmonic field in the STM junction is also important in ultrasensitive and ultrahigh resolution optical spectromicroscopy such as tip-enhanced Raman[22, 23, 24] and photoluminescence spectroscopy.[25, 26] Recent low-temperature tip-enhanced Raman spectroscopy showed that the exceptional sensitivity can be obtained at a very short gap distance between the tip apex and the molecule (below a few Å).[23, 24] Moreover, it was found that further enhancement of Raman scattering occurs by forming an atomic-point contact between the tip apex and the surface.[27] Interestingly, this enhancement is operative even for non-plasmonic substrate and we observed the surface phonons of the Si(111)-7×7 reconstructed surface.[28] These findings suggests atomic contacts have a substantial impact on light scattering in a plasmonic picocavity.

Here we investigate inelastic light scattering (ILS) from a plasmonic picocavity in the cryogenic STM junction which consists of an Ag tip and a single Ag adatom on the Ag(111) surface. The exceptional stability of the STM junction allows to examine different electron transport regimes of a picocavity, namely in the tunneling and quantum point contact (QPC) regimes. Previous studies addressed a correlation between surface-enhanced Raman scattering and conductive properties in molecular junctions using mechanical break junctions[29] and plasmonic nanodimers,[30] providing a microscopic insight into the Raman process in current-carrying molecular junctions. Our approach using low-temperature STM can image and manipulate individual atoms in an atomically well-defined environment. This will be an advantage to examine fundamental principles of ångström-scale optomechanics. In this paper, first we show how the ILS spectra evolve during reversible single-atom QPC formation in the STM junction. Tip-enhanced Raman scattering (TERS) from the vibration localized at the single Ag adatom is used to trace the evolution of the structure and plasmonic field. This measurement reveals that the TERS intensity and peak position exhibit a sharp change upon the QPC formation, implying a concurrent change of the atomic-level structure and plasmonic properties. Second, we explore the atomic-level structure of the STM junction and its plasmonic properties using time-dependent density functional theory (TD-DFT). The simulations reveal that atomic-scale structural relaxation at the single-atom QPC results in a discrete change of the plasmonic field strength, volume, and distribution as well as the vibration mode localized at the single atom. Third, we demonstrate that strong ILS is a characteristic feature of QPC by continuously forming, breaking, and reforming the atomic contact in the STM junction. Lastly, we show how the plasmonic resonance evolves throughout the non-tunneling, tunneling, and QPC regimes, which suggests a transition from capacitively-coupled to conductive-coupled plasmons.

**Results/Discussion.** We study the picocavity that consists of an Ag tip and a single Ag adatom on the Ag(111) surface at 10 K (**Fig. 1a**). The junction is illuminated by 633-nm laser ($\lambda_{\text{ext}}$) that excites the LSPR in the junction. The ILS ($\lambda_{\text{ILS}}$) are collected in a back-scattering geometry. The left panel of **Fig. 1b** shows an STM image of the Ag adatom created by single-atom transfer from the Ag tip.[31] The right panel of **Fig. 1b** shows the ILS spectra recoded over the adatom, where a characteristic peak is observed at 170 cm$^{-1}$ over the adatom. This peak is localized at the adatom and can be assigned to the bouncing mode of the adatom (*cf.* **Fig. 2**). A broad spectral feature at <160 cm$^{-1}$ remains over the flat surface where the narrow peak at 170 cm$^{-1}$ disappears. We assign the broad feature to phonons of the STM tip. A similar feature is observed in surface-enhanced Raman spectroscopy, and it is attributed to Raman scattering by phonons of a bulk Ag.[13] **Figure 1c** displays a waterfall plot of the ILS spectra recorded as a function of the vertical displacement between the tip and the surface ($\Delta z$). The vertical and horizontal axis corresponds to $\Delta z$ and the frequency shift ($\Delta v$) from the excitation wavelength (633 nm), respectively, and the color scale represents the ILS intensity ($I_{\text{ILS}}$). The left panel of **Fig. 1c** shows the STM current ($j_{\text{STM}}$) simultaneously recorded with the ILS spectra. The zero-point of the $\Delta z$ corresponds to an STM set-point given by $j_{\text{STM}}$=0.1 nA at a bias voltage ($V_{\text{bias}}$) of 1 V. During the measurement the $V_{\text{bias}}$ was set nominally to zero in the application software of the STM, but there was an offset bias of ~1.8 mV and a small photovoltage occurs under illumination. The photovoltage in the STM junction depends on the microscopic tip structure and is in the range up to a few tens of mV. In the tunneling regime, the $j_{\text{STM}}$ exponentially increases and, at a given distance, it exhibits a "jump-to-contact" behavior that is commonly observed for QPC formation in the STM junction.[32] The conductance right at the QPC formation is slightly smaller than the quantum conductance ($G_0$=7.748×10$^{-5}$ S), but it approaches asymptotically to $G_0$ when the QPC is squeezed.[37] **Figure 1c** reveals that strong enhancement of the $I_{\text{ILS}}$ occurs near the QPC. Because the $j_{\text{STM}}$–$\Delta z$ curve and the ILS spectra are almost symmetric with respect to the turning point of the approach and retraction ($\Delta z=-3$ Å), the measuring process is reversible, thus the structure of the STM junction is not significantly altered during the measurement. The results in **Fig. 1c** were reproducible for other adatoms and under different tip conditions (Supporting Information, **Fig. S1**). **Figures 1d** and **1e** show the $\Delta z$-dependence of the peak position ($\Delta v_{adatom}$) and the intensity ($I_{adatom}$) of the localized mode at the adatom. **Figure 1f** shows the simultaneously recorded $j_{STM}$. The peak position is gradually red-shifted in the tunneling regime, and upon the QPC formation it exhibits a rapidly red-shift. Then, the vibration frequency becomes almost constant as the junction is further squeezed in the QPC regime. The $I_{adatom}$ exponentially increases with decreasing the $\Delta z$ in the tunneling regime, indicating enhancement of the plasmonic field in the junction. However, the intensity of the adatom mode and the tip phonons suddenly drops upon the QPC formation.

To understand the $\Delta z$-dependence of tip-enhanced ILS spectra, we performed ground-state DFT simulations, considering two different geometries for the tip as shown in **Figs. 2a** and **2b**. Tip A has a single atom at the apex, whereas the apex atom is removed in tip B and its apex is composed of three atoms. **Figures 2c** and **2d** show the interatomic distances between the tip-apex atoms and the adatom ($d_{opt}$) as a function of the initial, unrelaxed vertical displacement ($\Delta z_{int}$). As evidenced by the structural relaxation curves (the markers in **Figs. 2c** and **2d**), tip B exhibits a jump-to-contact behavior resulting from sudden adatom displacement towards the tip apex, whereas this feature lacks in tip A. This difference is explained by the increased interatomic electronic density and larger binding energy of tip B compared to those of tip A (Supporting Information, **Fig. S2**).

The jump-to-contact in the structure relaxation observed for tip B (**Fig. 2d**) also results in an abrupt change of the vibration mode. We investigated the vibration modes localized at the adatom in the vicinity of the QPC for both tip structures by normal mode analysis. The mode localized at the adatom has a much higher frequency than the rest of the modes involving the adatom motion (Supporting Information, **Fig. S3**). As seen in **Figs. 2e** and **2f**, before point contact formation ($\Delta z_{int}>4$ Å), the localized mode exhibits a frequency of ~168 cm$^{-1}$ which is in good agreement with the observed narrow peak at ~170 cm$^{-1}$ in experiment (**Fig. 1**). This mode corresponds to a bouncing motion of the adatom (the right panel in **Figs. 2g** and **2h**, labeled by 4.6 Å). As the gap distance is reduced, a gradual and sudden decrease of the vibration frequency occurs in tip A and B, respectively (**Figs. 2e** and **2f**). The gradual change in tip A results from the localized interaction between the adatom and the very apex atom on the tip. However, the sudden change in tip B involves the mode transition from the adatom bouncing motion to a different vibration mode, resulting from chemical bonding between the adatom and the tip apex (as discussed in detail later).

In experiment, upon the QPC formation, we observed a rapid change in $I_{adatom}$ and $j_{STM}$ occurring within <0.1 Å (**Figs. 1e** and **1f**) and $\Delta\nu_{adatom}$ of 20–30 cm$^{-1}$ within 0.2–0.3 Å (**Fig. 1d** and see also **Fig. S1**). This behavior suggests that a jump-to-contact behavior in tip B is more consistent with experiment. A more moderate change in $\Delta\nu_{adatom}$ than that of $I_{adatom}$ will be explained by softening of the vibration mode through the van der Waals interaction between the adatom and the tip apex. In addition, the adatom may move forward and backward between the tip apex and surface near the QPC due to structural fluctuation at finite temperature, which would smooth slightly the abrupt change of the vibration frequency in tip B. Therefore, we focus on discussing the results for tip B. Previous experiments using atomic-force microscopy (AFM) also suggest the presence of tips with multiple apex atoms.[33a, 33b, 34, 35] It should be noted that there should exist other tip structures that better fit to the experiment. For instance, a tilted geometry of tip A could be a possibility (see Supporting Information, **Fig. S4**). This geometry exhibits a similar relaxation curve with tip B and is likely to involve the transition of the vibration mode upon the point contact formation. However, it is not realistic to find the exact tip geometry to have a perfect correspondence between theory and experiment, hence we focus on discussion of the correlation between the atomic-scale structure change and the plasmonic field in the STM picocavity. Determining exact tip apex shape will be possible using AFM in combination with DFT simulations,[38, 39, 40] so simultaneous STM/AFM/TERS measurements may be able to address this question in detail.

In the tip B configuration, a weak red-shift of the vibration frequency before QPC formation is caused by subtle pulling of the adatom by the tip due to the attractive interaction, which lowers the spring constant of the adatom bouncing mode. As mentioned above, the sudden drop of the vibration frequency at the QPC occurs by transition of the vibration mode. Before QPC, it is the adatom bouncing motion with negligible displacement of the tip atoms, whereas, after QPC, it becomes a coupled adatom–tip mode in which displacement of the tip atoms contributes significantly (the left panel in **Fig. 2h**). Because the coupled adatom–tip mode is more collective, its frequency is smaller than that of the adatom bouncing mode, similarly to what is observed in metal nanoclusters.[36] Because of a strong interaction between the adatom and the tip apex, the change in the vibration frequency upon the QPC formation is expected to be more sensitive to the tip structure, in agreement with the experiment (Supporting Information, **Fig. S1**).

The sudden structural relaxation in tip B leads to a discrete change of the plasmonic field in the STM picocavity. **Figure 3** displays the simulated spatial distribution of the longitudinal near field at 633-nm excitation. In the graph, we plot the maximum field in the gap ($|E_{gap}|$)$_{max}$ and the effective mode volume that is given by $V_{eff} = \int \frac{|E_{ind}(x,y,z)|^2}{|E_{ind,max}|^2} dV$.[16] When the cavity size enters the tunneling regime, charge-transfer plasmon gradually appears and capacitively-coupled plasmon, which is called bonding dimer plasmon in nanoparticle dimers, is quenched.[6, 7, 8, 16] This is evidenced here by the exponential dependence of the gap current on the distance above the QPC, and saturation below QPC (see Supporting Information, **Fig. S5**). As shown in **Fig. 3**, before QPC formation the field enhancement increases as the gap distance becomes smaller. Upon QPC formation, a sudden decrease and increase of ($E_{gap}$)$_{max}$ and $V_{eff}$, respectively, occurs, which is caused by the increase of conduction when the electronic densities of tip and substrate overlap[37] (see **Fig. S2** and **Fig. S5**). Therefore, the sudden change in near-field distribution as a function of tip position (see the field distribution at 4.0 and 3.8 Å in **Fig. 3**) results from "jump-to-contact" of the adatom to the tip apex. This field evolution is consistent with the observed $I_{adatom}$ (**Fig. 1e**) and highlights the need to consider single-atom motion to accurately predict the optical properties of plasmonic picocavities,[8] which could affect optical properties in ultranarrow plasmonic gaps.[38]

We demonstrate that strong ILS is a characteristic feature of the QPC. **Figure 4** shows the $\Delta z$-dependence of the scattering spectra recorded over the bare Ag(111) surface. The $j_{STM}$–$\Delta z$ curve again exhibits a jump-to-contact behavior corresponding to the QPC formation between the tip and the surface.[36, 37] By comparing with the simultaneously recorded STM current, it is clear that strong ILS appears in the vicinity of the QPC (region II). The narrow and broad spectral features at lower and higher frequency regimes are assigned, respectively, to Raman scattering by tip phonons and plasmonic background resulting either from electronic Raman scattering or photoluminescence.[39] However, these intense spectral features are completely quenched when a sudden decrease of the $j_{STM}$ occurs. This change in the $j_{STM}$ should correspond to a large structural change in the junction, most probably the formation of "multi-atom contact" (region III). Such a multi-atom contact in the STM junction was directly observed in transmission electron microscopy.[40] The drop of the $j_{STM}$ may be caused by a change of the photovoltage in the STM junction. We found that the photovoltage depends on the junction structure, but it is uncontrollable. In the case of **Fig. 4**, probably a decrease of the photovoltage occurred when the multi-atom contact is formed. However, strong ILS scattering resumes when the tip is retracted from the surface (region IV). In this region, the $j_{STM}$ becomes the same as that near the contact during the approach. Finally, the ILS signal disappears when the $j_{STM}$ drops, corresponding to breaking of the atomic-scale contact (region V). A similar atomic-scale contact (QPC) formation and breaking has been commonly observed in mechanical break junction experiments.[41, 42] A sequential process including formation of single- and multi-atom contacts, reformation and breaking of the single-atom contact, is reproduced in the simulations for tip B (see Supporting Information, **Movie 1** and **2**).

Finally, we examine the plasmonic response of the STM junction throughout the non-tunneling, tunneling, and QPC regimes. **Figure 5a** shows the ILS spectra recorded as a function of $\Delta z$. The tip is moved toward a single Ag adatom on the surface and there is a broad continuum reflecting the LSPR of the STM junction. **Figures 5b** and **5c** plot the intensity ($I_{LSPR}$) and the peak position ($\lambda_{LSPR}$) of the LSPR, respectively. $I_{LSPR}$ increases with decreasing $\Delta z$ in the non-tunneling regime, but it decreases as the $\Delta z$ enters the tunneling regime, where $\lambda_{LSPR}$ is red-shifted with decreasing $\Delta z$. This is a typical behavior of a capacitively-coupled plasmon mode.[6, 7, 8, 16] In nanoparticle dimers, it is well established that a red-shift of the resonance frequency arises from dipolar coupling and a decrease of the field intensity results from incremental quenching of a capacitively-coupled plasmons due to electron tunneling between nanoparticles. However, as seen in **Fig. 5b**, $I_{LSPR}$ increases near the QPC regime and $I_{LSPR}$ eventually saturates after QPC formation. This could be attributed to a contribution from conductively-coupled plasmons. The observed $\lambda_{LSPR}$ appears to be slightly red-shifted after QPC. The variation of the resonance frequency of charge-transfer plasmons depends on the cavity geometry, particularly on the conductivity of nanojunctions. While a larger conductivity in the cavity leads to a blue-shift of the resonance, a red-shift also could happen when the conductivity is small, which is expected for atomic-scale contacts.[43]

**Conclusions.** Using low-temperature optical STM and DFT simulations, we studied how the structural, electronic and plasmonic properties evolve near a single-atom QPC in the STM junction. The precise control of the STM junction enabled to continuously monitor the evolution of ILS during the transition from the tunneling to QPC regimes. We showed that strong enhancement of ILS occurs in the vicinity of the QPC, allowing to observe the vibration mode localized at a single Ag adatom. The QPC formation was manifested as a jump-to-contact behavior in the gap-distance dependence of the simultaneously recorded STM current, vibration frequency shift and Raman intensity of the localized mode at the single Ag adatom on the Ag(111) surface. We simulated the atomic-level structure, relaxation, the electronic conduction and the vibration frequency around the QPC by the DFT calculations. The simulations reveal a possible structure of the STM picocavity which reproduces the observed jump-to-contact behavior in experiment. Furthermore, we performed time-dependent DFT simulations to examine the field enhancement and confinement in the STM picocavity. The simulations revealed that atomic-level structure relaxation near the QPC plays a crucial role and the jump-to-contact of a single atom in the picocavity leads to a discrete change of the electric field strength, volume, and distribution. The increase of the field strength near the QPC regime is consistent with recent ultrahigh resolution TERS, which is measured in the "deep" tunneling regime.[22, 23] We would speculate Raman scattering in a current-carrying picocavity. Although Raman scattering is typically driven by direct interaction with an electric field, in the presence of charge transfer plasmons the current-driven Raman process may concurrently contribute as discussed in Ref. 44. In the point contact regime, the simulated local field is decreased (**Fig. 3**) but the flow of electrons is facilitated. Under these conditions, the current-driven Raman scattering might contribute substantially. It should also be noted that a change of the ILS intensity upon the QPC formation should be affected by the charge transfer mechanism, which is known as one of "chemical" enhancement effects leading to a change in the polarizability.[45] This mechanism has been considered to be responsible for additional enhancement of Raman scattering in current-carrying molecular junctions.[46, 47] In-depth understanding of light scattering in a plasmonic picocavity is a key for the further development of ultrasensitive and ultrahigh resolution tip-enhanced spectroscopy. In this sense, accurate description of atomic-scale light–matter interactions considering the detailed atomistic features in plasmonic nanostrucutres as well as quantum mechanical effects under electromagnetic field perturbation fully cupled to Maxwell's equations, could provide valuable insight of light–matter interactions in picocavities.[48] Furthermore, plasmonic picocavities are of significant importance in nanoscale optomechanics and optoelectronics which rely on atomic-scale structural modification for controlling optical properties.[49] Optical STM in combination with atom manipulation will offer a capability to study light–matter interaction at plasmonic picocavities in great detail and to will be a valuable tool for ångström-optics.

**Methods/Experimental**

**Sample and tip preparation.** All experiments were performed in an ultra-high vacuum (UHV) chamber (base pressure <5×10$^{-10}$ mbar). The Ag(111) surface was cleaned by repeated cycles of Ar$^+$ sputtering and annealing up to 670 K. We used an Ag tip polished by focused ion beam which can produce a plasmonically active tip in a highly reproducible manner.[50] The tip was sputtered by Ar$^+$ in the UHV chamber before measurement. The tip was also poked into the clean Ag(111) surface in a controlled manner. Therefore, the tip apex should be covered by Ag atoms.

**STM measurement.** We used a low-temperature STM from UNISOKU Ltd. (modified USM-1400) that is operated with Nanonis SPM Controller from SPECS GmbH. The bias voltage ($V_{bias}$) was applied to the sample, and the tip was grounded. The tunneling current ($I_t$) was collected from the tip.

**Raman measurement.** The excitation laser was focused to the STM junction with an *in-situ* Ag-coated parabolic mirror (numerical aperture of ~0.6) mounted on the cold STM stage. In the Raman measurements we used HeNe laser (633 nm). The incident beam is linearly polarized along the tip axis (p-polarization). The scattered photons are collected by the same parabolic mirror and detected outside of the UHV chamber with a grating spectrometer (AndorShamrock 303i). The parabolic mirror was precisely aligned using piezo motors (Attocube GmbH), which allow three translational and two rotational motions.

**DFT and TDDFT calculations.** The computational model consists of the Ag tip mimicked by a pyramidal cluster and the Ag(111) slab with a single adatom placed on the fcc site. Geometry relaxations and normal mode calculations of all structures (surface slab, tip and all the junctions with different gap distances) were performed at the DFT level using the all-electron FHI-aims code[51] at the generalized gradient approximation by Perdew-Burke-Ernzerhof (PBE).[52] Periodic boundary conditions were used considering a uniform 4×4×1 k-point grid centered around the Γ-point. The top-most layer of atoms of the tip and the bottom-most layer of the surface were fixed during the geometry optimizations. Structures were relaxed with a trust radius method enhanced version of the BFGS optimization algorithm until a maximum residual force per atom of 0.01 eV/Å was reached. For the normal-mode calculations, the atoms of the two bottom layers of the surface slab and the top-most layer of the tip were considered to be fixed. The modes with highest relative displacement of the adatom in the *z*-direction were considered for the analysis of the present work.

Time-domain simulations under laser illumination of the systems were carried out using the Octopus package[53] in the adiabatic local density approximation to describe exchange-correlation and a norm-conserving HGH pseudopotential to account for core electrons.[54] Numerical representation on a real-space grid was done considering a 0.35-bohr spacing and a time step of 1.57 as was used for time-propagation until a total simulated time of 30.0 fs was reached. During the dynamics the atomic positions were fixed, as the normal modes of interest have much longer periods than the maximum time of the simulation. The incoming laser was considered as a continous-wave of 633 nm wavelength (period of 2.11 fs) and a peak field strength of 0.26 V/Å (consistency with a weaker field strength of 0.026 V/Å was checked and is shown in the Supporting Information, **Figs. S6** and **S7**), while the time-dependent longitudinal near-field is defined as the gradient of the time-evolving Hartree potential difference with respect to the ground state. For the integration of the near-field needed to calculate the effective mode volume, a box size of 26 Å×26 Å×30 Å was considered.

## ASSOCIATED CONTENT

**Supporting Information**

The Supporting Information is available free of charge on the ACS Publications website at DOI: xxxx.

Content: Reproducibility of the ILS spectra recorded over a single adatom, Analysis of the ground state electronic density and the tip-adatom binding energy, Normal modes with the largest adatom displacement, Relaxation curve for the tilted tip A configuration, Simulation of the tunneling current in the STM junction, Simulations with a weaker field strength, Simulation of formation, collapse and regeneration of the QPC.

## AUTHOR INFORMATION

**Author notes**

**Corresponding author**

Correspondence to Takashi Kumagai.

kuma@ims.ac.jp

**Ethics declarations**

**Competing interests**

The authors declare no competing financial interests.

**Acknowledgements**

The authors thank Adnan Hammud for providing the Ag tips fabricated by using focused ion beam and Yair Litman for the geometries of the Ag tips used for the simulations. T.K. acknowledges the support by JST FOREST Program (Grant Number JPMJFR201J, Japan) and the Grants-in-Aid for Scientific Research (JSPS KAKENHI Grant Number 19K24684) from the Ministry of Education, Culture, Sports, Science, and Technology of Japan. F.B. acknowledges financial support from the European Union's Horizon 2020 research and innovation programme under the Marie Skłodowska-Curie grant agreement no. 895747 (NanoLightQD).

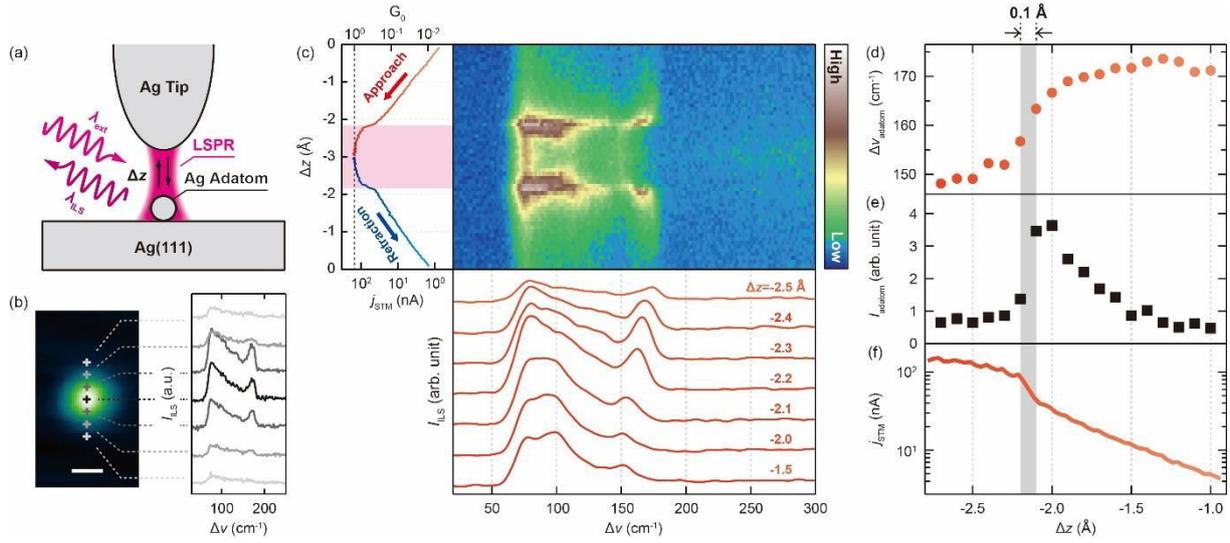

**Figure 1. ILS spectra measured for the STM picocavity.** (a) Schematic of the measurement. (b) STM image of a single Ag adatom on the Ag(111) surface at 10 K ($V_{bias}$=1 V, $j_{STM}$=0.1 nA, scale bar=5 Å) and the ILS spectra recorded at the marked positions with a 2 Å interval ($V_{bias}$=0 V, $\lambda_{ext}$=633 nm, $P_{ext}$=0.37 mW/μm$^2$). (c) (Top right) Waterfall plot of the ILS spectra recorded during tip approach and retraction ($V_{bias}$=0 V, $\lambda_{ext}$=633 nm, $P_{ext}$=0.37 mW/μm$^2$). (Top left) Simultaneously recorded $j_{STM}$–$\Delta z$ curve. Although the $V_{bias}$ is nominally set to zero, there is a voltage offset of ~1.8 mV and the photovoltage occurs under illumination. The pink shaded area indicates the QPC regime. (Bottom) ILS spectra at different $\Delta z$ during tip approach. $\Delta z$=0 corresponds to an STM set-point of $j_{STM}$=0.1 nA and $V_{bias}$=1 V. (d) and (e) $\Delta z$ dependence of the $\Delta v_{adatom}$ and $I_{adatom}$ in approach (f) Simultaneously recorded $j_{STM}$–$\Delta z$ curve.

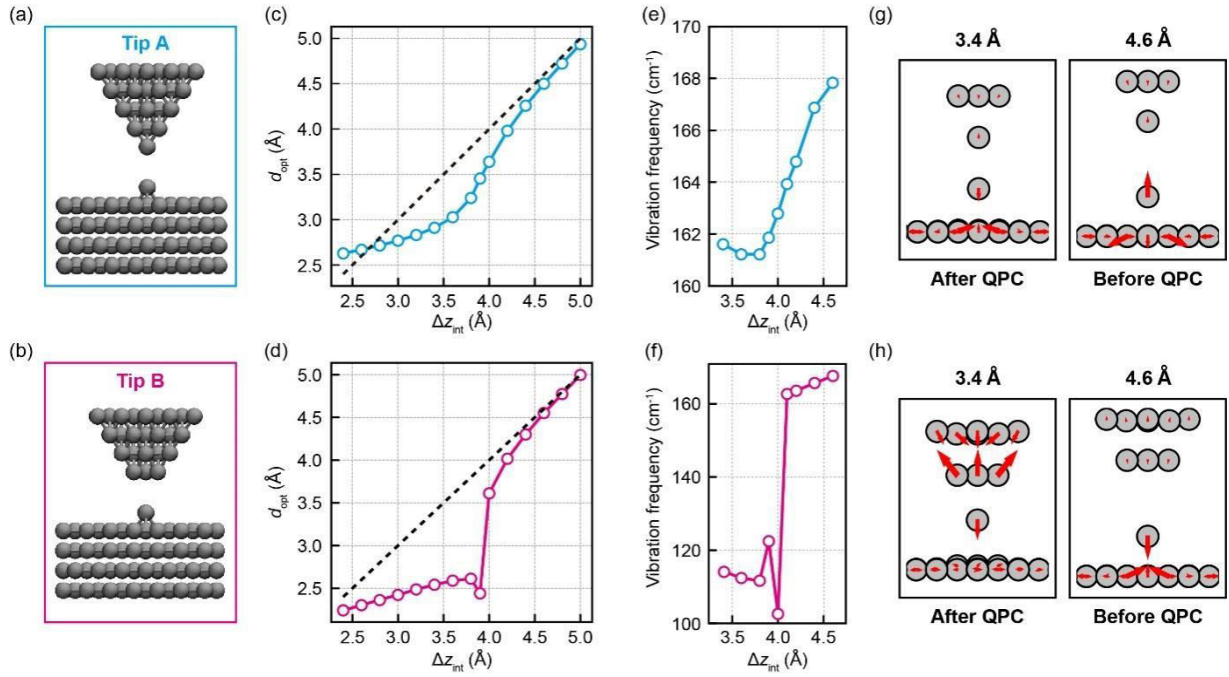

**Figure 2. Relaxed gap distance, peak currents and vibration frequencies.** (a) and (b) Model structures of the simulations. (c) and (d) Interatomic distances between the tip-apex atoms and the adatom nuclei as a function of the initial, unrelaxed vertical displacement, for tip A and tip B. (e) and (f) Frequency of the normal modes with the highest displacement of the adatom plotted as a function of $\Delta z_{int}$. (g) and (h) Displacement of the atoms along the normal modes with the highest displacement of the adatom at different distances before and after QPC formation.

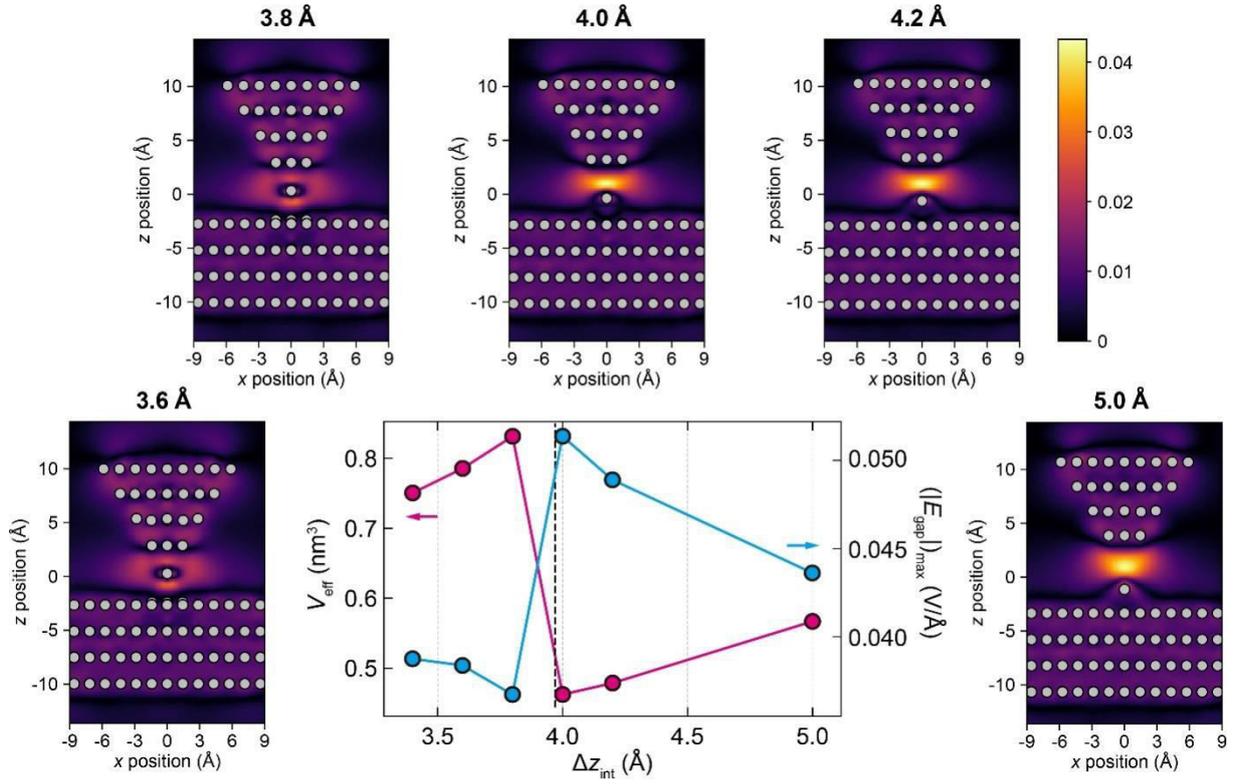

**Figure 3. Simulation of plasmonic field in the STM picocavity.** Snapshots of the *z*-component of the longitudinal field is plotted for several tip-adatom relative distances. The field is calculated as the Fourier transform of the time-dependent near field at the driving frequency (1.96 eV). The intensity color bar is in units of V/Å. The plot shows the effective mode volume and the maximum of the near field in the gap as a function of the vertical displacement of the tip.

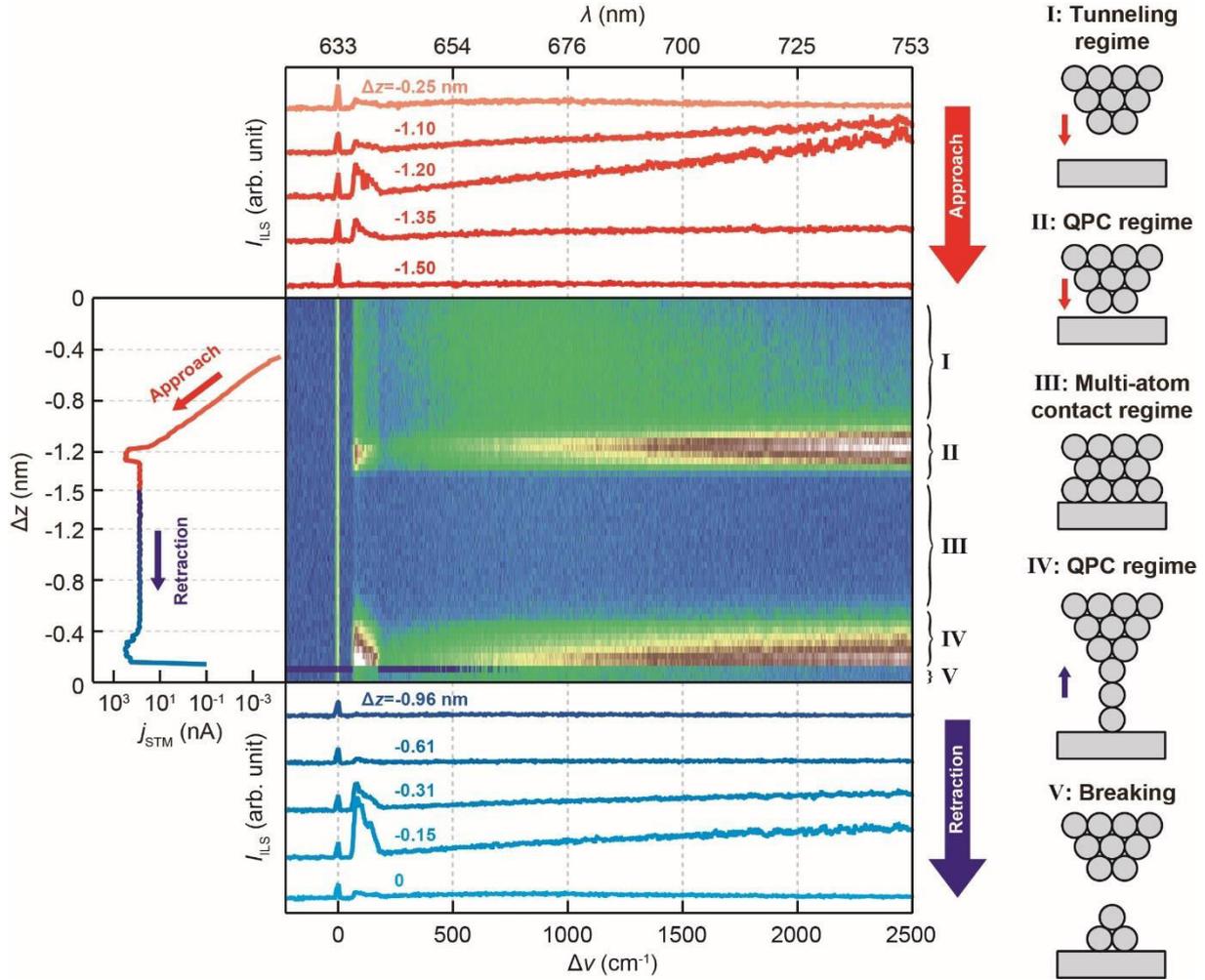

**Figure 4. ILS spectra measured for formation, breaking and reformation of QPC.** (Center) Waterfall plot of the ILS spectra recorded during tip approach and retraction measured over the bare Ag(111) surface (10 K, $V_{bias}$=0 V, $\lambda_{ext}$=633 nm, $P_{ext}$=0.37 mW/μm$^2$). $\Delta z$=0 corresponds to the distance of 0.5 nm away from an STM set-point of $j_{STM}$=0.1 nA and $V_{bias}$=1 V. Although the $V_{bias}$ is nominally set to zero, there is a voltage offset of ~1.8 mV and the photovoltage occurs under illumination. (Left) Simultaneously recorded $j_{STM}$–$\Delta z$ curve. (Top and bottom) ILS spectra at different $\Delta z$ during approach and retraction, respectively. (Right) The schematic models of the junction in the different regimes. I: tunneling regime, II: QPC formation, III: multi-atom contact formation, IV: QPC (single-atom wire) reformation, V: breaking of the QPC.

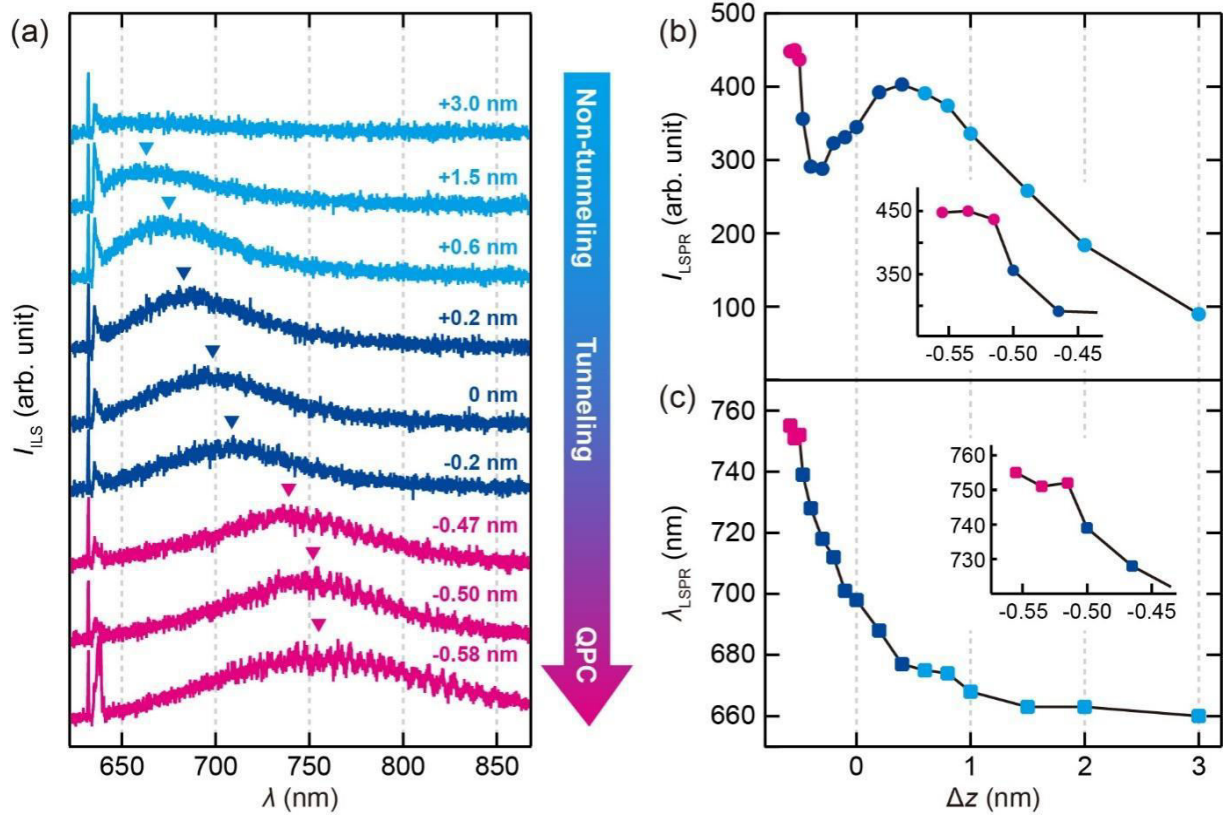

**Figure 5. Plasmonic response in the STM junction throughout non-tunneling, tunneling and QPC regimes.** (a) $\Delta z$-dependence of the ILS spectra when the Ag tip is moved vertically over the Ag adatom on the Ag(111) surface (10 K, $V_{bias}$=0 V, $\lambda_{ext}$=633 nm, $P_{ext}$=0.37 mW/μm$^2$). $\Delta z$=0 corresponds to an STM set-point of $j_{STM}$=0.1 nA and $V_{bias}$=1 V. Although the $V_{bias}$ is nominally set to zero, there is a voltage offset of ~1.8 mV and the photovoltage occurs under illumination. The triangles mark the peak position. (b) and (c) Peak intensity and position of the spectra as function of gap distance, respectively.